\documentclass{cimento}
\title{Mass neutrino flavor evolution in  spacetime with torsion}
\author{C. M. Zhang}
\instlist{
\inst  Department of Physics, Hebei University of Technology,\\
Tianjin-300130, P.R. China\\

\inst  Instituto de F\'{\i}sica Te\'orica,  
Universidade Estadual Paulista\\
Rua Pamplona 145, 01405-900\, S\~ao Paulo, Brazil}

\PACSes{\PACSit{04.50, 04.90}{}}

\begin{document}

\maketitle

\def\be{\begin{equation}}  
\def\ee{\end{equation}}
\def\ba{\begin{eqnarray}}
\def\ea{\end{eqnarray}}
\def\nn{\nonumber}
\def\nonu{\nonumber}
\def\ep{\epsilon}
\def\ga{\gamma}
\def\Ga{\Gamma}
\def\la{\lambda}
\def\si{\sigma}
\def\al{\alpha}
\def\m{\mu}
\def\pa{\partial}
\def\de{\delta}   
\def\De{\Delta}
\def\rsr{{r_{s}\over r}}
\def\rrs{{r\over r_{s}}}
\def\rs2r{{r_{s}\over 2r}}
\def\l2r2{{l^{2}\over r^{2}}}
\def\rsa{{r_{s}\over a}}
\def\rsb{{r_{s}\over b}}
\def\rsro{{r_{s}\over r_{o}}}
\def\rss{r_{s}}
\def\a2{{l^{2}\over a^{2}}}
\def\b2{{l^{2}\over b^{2}}}
\def\op{\oplus}

\begin{abstract}
\noindent
In the framework of the spacetime with torsion,   we 
obtain the flavor evolution equation of the mass neutrino oscillation in vacuum. 
A comparison with the result of  general
relativity case, it shows that the  flavor evolutionary  equations in Riemann
spacetime  
and Weitzenb\"ock spacetimes are   equivalent in the spherical symmetric  
Schwarzschild spacetime, but turns out to 
 be different in the case of  the axial symmetry. 
\end{abstract}

\section{Introduction}
On account of the  Super-Kamiokande atmospheric neutrino experiment, which
confirmed a nonvanishing mass for the neutrino~\cite{sk}, the
neutrino oscillation problem became an even hotter topic in high energy
physics, both from the experimental and from the theoretical points of
view~\cite{zub98,bil98,pak98}. As a natural extension to this problem,
 many authors 
consider the neutrino oscillation in the presence of gravitation,
that
is, in a curved spacetime. This means to extend the physics related to the
neutrino oscillation in a Minkowski spacetime with Lorentz invariance, to
a   
Riemannian spacetime with the usual invariance under general coordinate
transformation.
 The gravitational effect on the neutrino oscillation has attracted much
attention
recently~\cite{wud91,ahl96,ahl97,gro96,ful96,bha96,for96,wud96,bru98} 
in the framework of general relativity, although
 a lot of problems concerning the understanding of the
gravitationally induced neutrino oscillation still persists. 
 However many alternative mechanisms have been proposed to account
for 
the gravitational effect on the neutrino oscillation; {\it e.g.} the
 equivalence principle and neutrino oscillation~\cite{gas88,man}. 
As a further theoretical exploration, 
more recently torsion induced neutrino oscillation in $U_{4}$ spacetime 
 \cite{heh76} with both curvature and torsion is also
proposed~\cite{sab81,ali99}. In this article, we extend the
neutrino oscillation problem into the spacetime with torsion but without 
curvature, {\it i.e.} in a Weitzenb\"ock spacetime $A_{4}$
\cite{hay79,ap,wei,per1}, in the framework of new general relativity
(NGR)\cite{hay79}.  

The paper is organized as follows. 
In Sec. II we briefly introduce the gravitational theory in
Weitzenb\"ock spacetime. In 
Sec. III we compare Dirac equations in Riemaniann spacetime and
 in Weitzenb\"ock spacetime,  and  in Sec. IV. we
derive    the evolutionary equation of the 
neutrino oscillation amplitude in Weitzenb\"ock spacetime. 
 We set $G = \hbar = c = 1$ throughout this article.

\section{A brief review of the NGR}

The new general relativity is a parallel tetrad gravitational theory, 
which  is formulated on a 
Weitzenb\"ock spacetime \cite{hay79,per1,zcm}. It is 
characterized by the vanishing curvature tensor and by the 
torsion tensor formed of four parallel tetrad vector fields. 
Namely, the 
 gravitational field appears as the nontrivial part of
the tetrad field. We will use the greek alphabet ($\mu$, $\nu$,
$\rho$,~$\cdots=1,2,3,4$) to denote tensor indices,  that is, indices
related to spacetime. The latin alphabet ($a$, $b$, $c$,~$\cdots=1,2,3,4$) 
will be  used to denote local Lorentz (or tangent space) indices. Of
course,
being of the same kind, tensor and local Lorentz indices can be changed
into each other with the use of the tetrad, denoted by $h^{a} {}_{\mu}$,
and supposed to satisfy
\be
h^{a}{}_{\mu} \; h_{a}{}^{\nu} = \delta_{\mu}{}^{\nu} \quad
; \quad h^{a}{}_{\mu} \; h_{b}{}^{\mu} =
\delta^{a}{}_{b} \; .
\label{orto}
\ee

As is known, curvature and torsion are properties of a
connection\cite{ap,per1},
and many  different connections may be defined on the same space. For
example, denoting by $\eta_{a b}$ the metric tensor of the tangent space,
a
nontrivial tetrad field can be used to define the riemannian metric
\be
g_{\mu \nu} = \eta_{a b} \; h^a{}_\mu \; h^b{}_\nu \; ,
\label{gmn}
\ee
in terms of which the Levi--Civita connection
\be
{\stackrel{\circ}{\Gamma}}{}^{\sigma}{}_{\mu \nu} = \frac{1}{2}
g^{\sigma \rho} \left[ \partial_{\mu} g_{\rho \nu} + \partial_{\nu}
g_{\rho \mu} - \partial_{\rho} g_{\mu \nu} \right]
\label{lci}
\ee
can be introduced. Its curvature
\be
{\stackrel{\circ}{R}}{}^{\theta}{}_{\rho \mu \nu} = \partial_\mu
{\stackrel{\circ}{\Gamma}}{}^{\theta}{}_{\rho \nu} +
{\stackrel{\circ}{\Gamma}}{}^{\theta}{}_{\sigma \mu}
\; {\stackrel{\circ}{\Gamma}}{}^{\sigma}{}_{\rho \nu} - (\mu
\leftrightarrow \nu) \; ,
\label{rbola}
\ee
according to general relativity, accounts exactly for the gravitational
interaction. Owing to the universality of gravitation, which means that
all particles feel ${\stackrel{\circ}{\Gamma}}{}^{\sigma}{}_{\mu \nu}$ the
same, it turns out possible to describe the gravitational interaction by
considering a Riemann spacetime with the curvature of the Levi--Civita
connection, in which all particles will follow geodesics. This is the
stage
set of Einstein's General Relativity, the gravitational interaction being
mimicked by a geometrization of spacetime.
   
On the other hand, a nontrivial tetrad field can also be used to define
the
linear Cartan connection
\be
\Gamma^{\sigma}{}_{\mu \nu} = h_a{}^\sigma \partial_\nu
h^a{}_\mu \; ,
\label{car}
\ee
with respect to which the tetrad is parallel:
\be
{\nabla}_\nu \; h^{a}{}_{\mu} \equiv
\partial_\nu h^{a}{}_{\mu} - \Gamma^{\rho}{}_{\mu \nu} \,
h^{a}{}_{\rho} = 0 \; .
\label{weitz}
\ee
For this reason, tetrad theories have received the name of
teleparallelism, or absolute parallelism. Plugging in Eqs.(\ref{gmn}) and
(\ref{lci}), we get
\be
{\Gamma}^{\sigma}{}_{\mu \nu} = {\stackrel{\circ}{\Gamma}}{}^
{\sigma}{}_{\mu \nu} + {K}^{\sigma}{}_{\mu \nu} \; ,
\label{rel}
\ee
where
\be
{K}^{\sigma}{}_{\mu \nu} = \frac{1}{2} \left[
T_{\mu}{}^{\sigma}{}_{\nu} + T_{\nu}{}^{\sigma}{}_{\mu} -
T^{\sigma}{}_{\mu \nu} \right]
\label{conto}
\ee
is the contorsion tensor, with
\be
T^\sigma{}_{\mu \nu} = \Gamma^{\sigma}{}_{\nu \mu} - \Gamma^
{\sigma}{}_{\mu \nu} \;
\label{tor}  
\ee

If now we try to introduce a spacetime with the
same properties of the Cartan connection $\Gamma^{\sigma}{}_{\nu \mu}$, we
end up with a Weitzenb\"ock spacetime~\cite{wei}, a space presenting
torsion, but no curvature. This spacetime is the stage set of the
teleparallel description of gravitation. Considering that local
Lorentz indices are raised and lowered with the Minkowski metric $\eta^{a
b}$, tensor indices on it will be raised and lowered with  the riemannian
metric $g_{\mu \nu} = \eta_{a b} \; h^a{}_\mu \;
h^b{}_\nu$~\cite{hay79}.
Universality of gravitation, in this case, means that all particles feel
$\Gamma^{\sigma}{}_{\nu \mu}$ the same, which in turn means that they will
also feel torsion the same.

{}From the above considerations, we can infer that the presence of a
nontrivial tetrad field induces both, a riemannian and a teleparallel
structures in spacetime. The first is related to the Levi--Civita
connection, a connection presenting curvature, but no torsion. The second
is related to the Cartan connection, a connection presenting torsion, but
no curvature. It is important to remark that both connections are defined
on the very same spacetime, a spacetime endowed with both a riemannian and
a teleparallel structures.

As already remarked, the curvature of the Cartan connection vanishes
identically:
\be
{R}^{\theta}{}_{\rho \mu \nu} = \partial_\mu
{\Gamma}^{\theta}{}_{\rho \nu} + {\Gamma}^{\theta}{}_{\sigma   
\mu} \; {\Gamma}^{\sigma}{}_{\rho \nu} - (\mu  \leftrightarrow \nu)
\equiv 0 \; .
\label{r}
\ee
Substituting ${\Gamma}^{\theta}{}_{\mu \nu}$ from
Eq.(\ref{rel}), we get
\be
{R}^{\theta}{}_{\rho \mu \nu} =
{\stackrel{\circ}{R}}{}^{\theta}{}_{\rho \mu \nu} +
Q^{\theta}{}_{\rho \mu \nu} \equiv 0 \; ,
\label{eq7}
\ee

The gravitational Lagrangian density in NGR is written in the form 

\be
{\cal L}_G= {\sqrt{-g}\over \kappa} \left[a_1(t^{\mu \nu \lambda}
t_{\mu \nu \lambda})+a_2(v^{\mu} v_{\mu})+a_3(a^{\mu}a_{\mu}) \right],
\ee
where $a_1$, $a_2$ and $a_3$ are
dimensionless parameters of the theory, 
\be
t_{\mu \nu \lambda}  =  {1 \over 2}
\left(T_{\mu \nu \lambda}
+T_{\nu \mu \lambda} \right)
+{1 \over 6} \left(g_{\lambda \mu} v_\nu
+g_{\lambda \nu} v_{\mu} \right)-{1 \over 3} g_{\mu \nu} v_\lambda, \\
\ee

\be
v_{\mu}  =  T^{\lambda}_{\lambda \mu}, 
\ee

\be
a_{\mu}  =  {1 \over 6}{\epsilon}_{\mu \nu \rho \sigma} T^{\nu \rho \sigma}
\ee
with ${\epsilon}_{\mu \nu \rho \sigma}$ being the completely antisymmetric
tensor normalized as ${\epsilon}_{0123}=\sqrt{-g}$.
By applying variational principle to the above Lagrangian, we get the field
equation:
\be
I^{\mu \nu}= {\kappa}T^{\mu \nu}, \,\,\, \kappa = 8 \pi,
\ee
with
\be
I^{\mu \nu}=2{\kappa}[{D}_\lambda F^{\mu \nu \lambda}+
v_\lambda F^{\mu \nu \lambda}+H^{\mu \nu}
-{1 \over 2} g^{\mu \nu}L_G],
\ee
where
\ba
F^{\mu \nu \lambda}  &= & {1 \over 2} h^{k \mu} {\partial L_G \over \partial
{h^k}_{\nu,\lambda}} \\\nonu
  &=  &{1 \over \kappa} \left[ a_1 \left(t^{\mu \nu \lambda}
-t^{\mu \lambda \nu} \right)+a_2 \left(g^{\mu \nu} v^\lambda
-g^{\mu \lambda} v^\nu \right)
-{a_3 \over 3} \epsilon^{\mu  \nu \lambda \rho} a_\rho \right]\\
& = & -F^{\mu \lambda \nu},\\
H^{\mu \nu} & = & T^{\rho \sigma \mu}
 {F_{\rho \sigma}}^\nu - {1 \over 2} T^{\nu \rho \sigma}
{F^\mu}_{\rho \sigma}=H^{\nu \mu},\\
{L_G}  & = & {{\cal L}_G \over \sqrt{-g}},\\
T^{\mu \nu} & = & {1 \over \sqrt{-g}} {\delta {\cal L}_M \over
\delta {h^k}_\nu} h^{k \mu}.
\ea
Here ${\cal L}_M$ denotes the Lagrangian density of material fields and
$T^{\mu \nu}$ is the material energy-momentum tensor which is
nonsymmetric in general.
In order to reproduce the correct Newtonian limit, we require the parameters
$a_1$ and $a_2$ to satisfy the condition
\be
a_1+4a_2+9a_1a_2=0,
\ee
called the Newtonian approximation condition,   which can be
solved to give
\be
a_1=-{1 \over 3(1-\epsilon)}, \quad
a_2={1 \over 3(1-4\epsilon)}
\ee
with $\epsilon$ being a dimensionless parameter. The comparison with
solar-system experiments shows that $\epsilon$ should be given by 
\be
\epsilon=-0.004 \pm0.004,
\ee

\section{Dirac equation in Weitzenb\"ock spacetime}

Previous to entering to our main point, we stress that the semiclassical
by approximated Dirac particle does not follow a  geodesic exactly.
 However the force aroused by the  spin and the curvature coupling
has little contribution to the geodesic deviation\cite{aud81}. So here we
take the neutrino as a spinless particle to go along the geodesic.
The gravitational effects on the spin are incorporated into Dirac
equation  through the ``spin
connection'' $\Gamma_{\mu}$ appearing in the Dirac equation
in curved spacetime \cite{sto79,ana,aud81},
which is constructed by means of the variation of the covariant Lagrangian
of the spinor field. In the parallel tetrad theory of  Hayashi and
Shirafuji \cite{hay79}, considering the covariant derivative of spinor 
 to coincide with the usual derivative, the Dirac Langrange density
$L_{D}$ 
is given by 

\be 
L_{D} = {1 \over 2}h^{\mu}_{k}[{\psi}\gamma^{k}\pa_{\mu}\bar{\psi} 
- \pa_{\mu}\bar{\psi}\gamma^{k}\psi ] - m \bar{\psi} \psi.
\ee
By taking variation with respect to $\bar{\psi}$,  the Dirac equation in
Weitzenb\"ock spacetime is given as

\be
\left[ \gamma^{a} h^{\mu}_{a} (\partial_{\mu} + \Gamma_{\mu})
        + m
        \right]
        \psi = 0,  \label{dirac2}
\ee
and the spin connection $\Gamma_{\mu}$ is  
\be
\Gamma_{\mu} = {1\over 2}v_{\mu}
\ee
where $v_{\mu}$ is the tetrad vector.

The spin connection $\Gamma_{\mu}$ is different from that of 
general relativity 
because the parallelism of vector in Weitzenb\"ock spacetime 
makes the covariant derivative of spinor to coincide the usual dirivative.
However, 
 the Lagrangian of Dirac equation in general relativity is constructed by 
the covariant derivative and its  explicit expression for
the spin connection $\Gamma_{\mu}$ is \cite{ful96}

\be
\Gamma_{\mu} = {1 \over 8}[\gamma^b, \gamma^c] h_{b}{}^{\nu}
        h_{c \nu;\mu}.
\ee
We must first simplify the Dirac matrix product in the spin
connection term. It can be shown that
\be
\gamma^a [\gamma^b, \gamma^c] = 2 \eta^{ab} \gamma^c -
        2 \eta^{ac} \gamma^b - 2i \epsilon^{dabc} \gamma_5
        \gamma_d,  \label{gammas}
\ee
where $\eta^{ab}$ is the metric of flat space and
$\epsilon^{abcd}$ is the (flat space) totally antisymmetric tensor,
with $\epsilon^{0123}= +1$. With Eq.(\ref{gammas}), the
contribution from the spin connection in general relativity is
\be
\Ga_{\mu}  = {1\over 2}v_{\mu}  - {3i \over 4}a_{\mu}\ga_{5},
\label{ga2}
\ee
where  
\be
a_{\mu}= {1\over 6}
\ep_{\mu\nu\la\si}T^{\nu\la\si}
\ee
$a_{\mu}$ is the
tetrad axial-vector represented the deviation of the axial symmetry 
from the spherical symmetry \cite{nit81}.

Or, in the spherical case, Schwarzschild spacetime, both 
Dirac equations in Riemaniann spacetime  and in Weitzenb\"ock spacetime
are
equivalent. 
The difference between them will appear if the spacetime includes the
axial symmetric components, Kerr spacetime for instance.

\section{Evolutionary  equation for the  neutrino oscillation amplitude}

As proceeded in ref.\cite{ful96}, in order to incorporate the
gravitational effect into the matter effect, we rewrite the spin
connection term as 
\be
\ga^a h^\m_a \Ga_\m =\ga^a h^\m_a (iA_{G\m} {\cal P}_L)
= \ga^a h_a^\m
\left\{
 i A_{G\m} \left[
  - \frac{1}{2\sqrt{-g}} \ga_5
 \right]   
\right\},
\ee
where ${\cal P}_L = - \frac{1}{2\sqrt{-g}} \ga_5$ is the left--handed
projection operator, and 

\be
A^{\mu}_G \equiv {2 i }{} (-g)^{1/2} \ga_5 v_{\mu}
\ee
In the  above equations,  $(-g)^{1/2} = [\rm det(g_{\mu\nu})]^{1/2}$. 
 Proceeding as in the discussion by Cardal and Fuller \cite{ful96}, 
we will borrow the three-momentum operator used in the 
neutrino oscillation, which can be calculated  from the mass shell
condition obtained by
iterating the Dirac equation
\begin{equation}
(P_{\mu}+ A_{G\mu }{\cal P}_L )
(P^{\mu} +  A^{\mu}_{G}{\cal P}_L ) = - M_{f}^2,
\end{equation}
where we have not included background matter effects.
 $M_f^2$ is the vacuum mass matrix in flavour basis
\be
M_f^2=U
\left( \begin{array}  {cc} m_1^2 &0 \\ 0&m_2^2
\end{array} \right) U^\dagger ,
\ee
where
\be
U=\left( \begin{array}  {cc} 
{\rm cos} \theta &{\rm sin} \theta \\ -{\rm sin} \theta & {\rm cos} \theta
\end{array} \right),
\ee
with $\theta$  the mixing angle between different eigenstates of
mass 
neutrinos. For relativistic neutrinos,
ignoring terms of ${\cal O}(A^2)$ and ${\cal O}(AM)$,
and remembering that we employ  a null tangent vector $n^{\mu}$,   
 which is defined as $n^{\m} = dx^{\m}/d\la$,  and \\
${x^{\m}}(\la) = \left[x^{0}(\la), x^{1}(\la), x^{2}(\la),
x^{3}(\la)\right]$, we find

\be
P_{\mu} n^{\mu} = - \left( M_{f}^{2}/2 + A_{G\mu}n^{\mu}\right). 
\ee\label{pn}
It is convenient to define a column vector of flavor
amplitudes. For example, for the mixing between $\nu_e$ and $\nu_{\tau}$,

\be\label{cl}
\chi(\la) \equiv \pmatrix{\langle \nu_e | \Psi(\la)
                \rangle \cr
                \langle \nu_{\tau} | \Psi(\la) \rangle}.
\ee
Eq.(\ref{cl}) can be written as a differential equation for
 the null world line  parameter $\la$,
\be
i {d\chi \over d\la} = \left(M_{f}^{2}/2 + A_{fG\mu}n^{\mu} \right)
\chi,
\label{dcds}
\ee
where the subscrip f denotes `` flavor basis".  Eq.(\ref{dcds})
can be integrated  to yield the
neutrino flavor evolution. Similar equations were obtained in
Refs.\cite{ful96,ali99} in Riemaniann spacetime and in $U_4$ 
spacetime respectively.

\section{Conclusion and discussion}

In this paper, we studied the evolution equation for the neutrino
oscillation amplitude in the framework of the new general relativity
\cite{hay79}. 
We find that our results will be equivalent to that of general relativity 
in the case of spherical symmetry, and the difference will occur 
when the axial tetrad vector is not zero.

\section*{Acknowledgments}

The author would like  to thank FAPESP-Brazil for
financial support, and  J.G. Pereira for
helpful discussions. Thanks are also due to the hospitalities from 
 S. Civaram, K. Hayashi,  F.W. Hehl and J.M. Nester when he
visited their research groups. 


\end{document}